\documentclass[reprint,amsmath,amssymb,aps,prl,superscriptaddress,]{revtex4-2}

\usepackage{subfiles}
\usepackage{graphicx}
\usepackage{dcolumn}
\usepackage{bm}
\usepackage{color}
\usepackage{braket}
\usepackage{hyperref}
\hypersetup{hidelinks}

\begin{document}
	

\title{All-optical frequency division on-chip using a single laser}

\author{Yun Zhao}
\thanks{These authors contributed equally to this work.}
\affiliation{Department of Applied Physics and Applied Mathematics, Columbia University, New York, NY 10027, USA}	
\affiliation{Department of Electrical Engineering, Columbia University, New York, NY 10027, USA}
\author{Jae K. Jang}
\thanks{These authors contributed equally to this work.}
\affiliation{Department of Applied Physics and Applied Mathematics, Columbia University, New York, NY 10027, USA}
\author{Karl J. McNulty}
\affiliation{Department of Electrical Engineering, Columbia University, New York, NY 10027, USA}
\author{Xingchen Ji}
\affiliation{Department of Electrical Engineering, Columbia University, New York, NY 10027, USA}
\author{Yoshitomo Okawachi}
\affiliation{Department of Applied Physics and Applied Mathematics, Columbia University, New York, NY 10027, USA}
\author{Michal Lipson}
\affiliation{Department of Electrical Engineering, Columbia University, New York, NY 10027, USA}
\affiliation{Department of Applied Physics and Applied Mathematics, Columbia University, New York, NY 10027, USA}
\author{Alexander L. Gaeta}
\email{a.gaeta@columbia.edu}
\affiliation{Department of Applied Physics and Applied Mathematics, Columbia University, New York, NY 10027, USA}
\affiliation{Department of Electrical Engineering, Columbia University, New York, NY 10027, USA}

\date{\today}

\begin{abstract}
	The generation of spectrally pure high-frequency microwave signals is a critical functionality in fundamental and applied sciences, including metrology and communications. The development of optical frequency combs has enabled the powerful technique of optical frequency division~(OFD) to produce microwave oscillations of the highest quality. The approaches for OFD demonstrated to date demand multiple lasers with space- and energy-consuming optical stabilization and electronic feedback components, resulting in device footprints incompatible with integration into a compact and robust photonic platform. Here, we demonstrate all-optical OFD on a single photonic chip driven with a single continuous-wave laser. We generate a dual-point frequency reference using the beat frequency of the signal and idler fields from a microresonator-based optical parametric oscillator~(OPO), which achieves high phase stability due to the inherently strong signal-idler frequency correlations. We implement OFD by optically injecting the signal and idler fields from the OPO to a Kerr-comb microresonator on the same chip. We show that the two distinct dynamical states of Kerr cavities can be passively synchronized, allowing broadband frequency locking of the comb state, which transfers the stability of the OPO frequencies to the repetition rate of the Kerr comb. A 630-fold phase-noise reduction is observed when the Kerr comb is synchronized to the OPO, which represents the lowest noise generated on the silicon-nitride platform. Our work demonstrates a simple, effective approach for performing OFD and provides a pathway toward chip-scale devices that can generate microwave frequencies comparable to the purest tones produced in metrological laboratories. This technology can significantly boost the further development of data communications and microwave sensing.
\end{abstract}

\maketitle

\begin{figure*}
	\centering
	\includegraphics{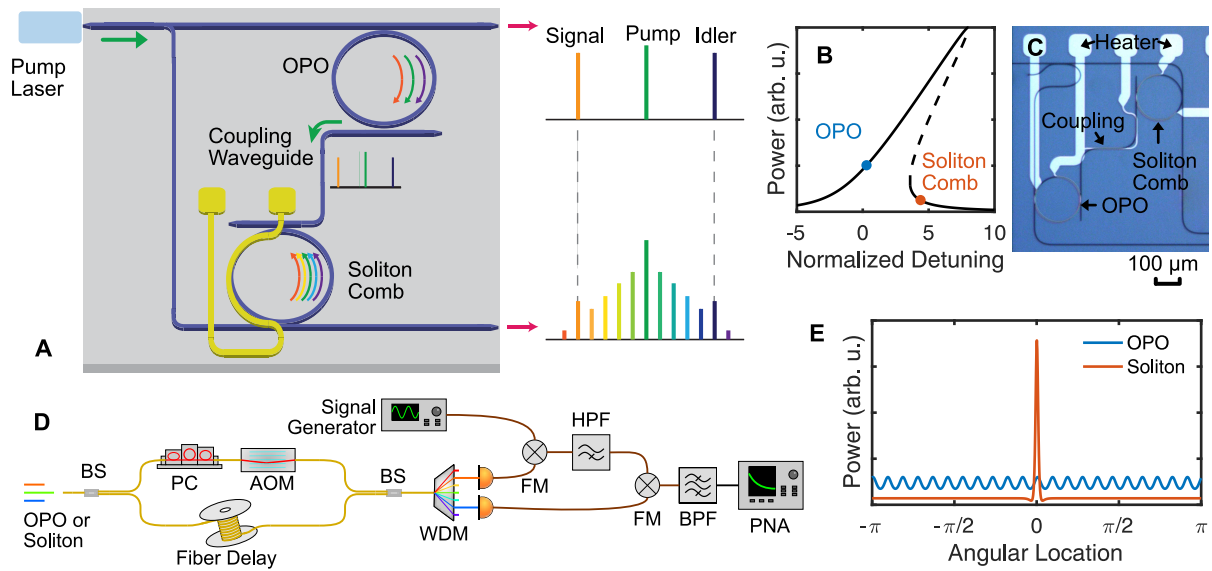}
	\caption{\textbf{Schematic of on-chip low-noise microwave generation via frequency division}. (A) A high-level schematic of our photonic integrated chip for all-optical OFD. The top microresonator operates in the OPO regime, resulting in the frequency-correlated signal-idler pair, which serves as a stable optical reference. The microresonator at the bottom operates in the Kerr-soliton regime, which has a repetition rate in the microwave domain. An evanescently coupled waveguide transfers the stability of the OPO reference to the comb via all-optical synchronization. The resonances of the microresonators are independently controlled through suitable application and modulation of electric currents on the integrated heaters (yellow trace). (B) Illustration of different dynamical branches of a Kerr cavity. The left (right) solid trace corresponds to the blue-detuned (red-detuned) branch, which supports the OPO (soliton) state. The dashed trace corresponds to a dynamically unstable branch. (C) The photonic chip used in the experiment. (D) Schematic of our delayed self-heterodyne setup for phase-noise characterization. BS is beam splitter, PC is polarization controller, AOM is acousto-optic modulator, WDM is wavelength-division multiplexer, FM is frequency mixer, HP is high pass filter, BP is bandpass filter, and PNA is phase noise analyzer. This system can measure the phase noise carried by the difference frequency of the blue and red components. (E) The waveforms of OPO and soliton inside the cavity. The OPO has shallow and dense oscillations, while the soliton has a sharp peak and a low CW background. In the experiment, 0.002\textperthousand~of the OPO power is coupled to the soliton comb.}
	\label{figScheme}
\end{figure*}

Stable microwave sources are an indispensable tool in today's electronic devices, which serve as clocks and information carriers in applications including communication, sensing, and data processing. Two key aspects that are actively pursued for microwave sources are high-frequency generation and ultra-low-noise operation, which can lead to higher information capacity in communication and higher sensitivity in metrology. Most commercially available chip-scale microwave sources rely on mechanical high-$Q$ oscillators with natural frequencies ranging from 10~kHz to 250~MHz. Synthesis of higher frequencies (for example, via phase-locked loops) from such low-frequency oscillators results in severe noise penalties due to frequency multiplication. To achieve the performance levels required for advanced applications such as metrology and high-speed data communications, various techniques have been developed for generating spectrally pure high-frequency microwaves, including electronic \cite{Ivanov_IEEE_96,Kinget_VCO_99,Razavi_IEEE_09,Rappapport_IEEE_11}, microelectromechanical \cite{vanBeek_Misc_11}, and optoelectronic \cite{Madjar_IEEE_06,Maleki_NatPhot_11,Li_Science_14} methods. In particular, high-quality optical oscillators are readily available at frequencies exceeding 100~THz, and by implementing a suitable frequency-down-conversion scheme such as optical frequency division~(OFD), microwave generation with a large noise suppression factor can be realized. This technique forms the basis of optical atomic clocks and yields the most precise microwave frequency generated to date \cite{Ludlow_RMP_15,Bothwell_Misc_19}. For low-phase-noise microwave generation, the narrow-linewidth-laser systems can be operated without cold atomic references. Furthermore, a two-point referencing scheme (Fig. \ref{figScheme}A) removes the need for an octave-spanning comb \cite{Li_Science_14,Tetsumoto_NatPhot_21}, further simplifying the setup. Nonetheless, such a system still requires multiple fast-tunable laser sources and multiple optical and electronic stabilization stages, resulting in a large, table-top level footprint \cite{Fortier_NatPhot_11,Li_Science_14,Tetsumoto_NatPhot_21}, which does not meet the compactness and robustness required by many sensing and communication applications. 

In this work, we propose and demonstrate an OFD scheme based on a photonic chip pumped with a single continuous wave~(CW) laser, which is capable of a large-bandwidth noise reduction without feedback control or high-speed pump-frequency modulation. As shown in Fig.~\ref{figScheme}A, the top microresonator operates as an optical parametric oscillation~(OPO), and the generated signal-idler pair produces a stable frequency reference due to phase-correlation-induced linewidth narrowing. In principle, the frequency separation between the OPO sidebands can reach an octave with proper dispersion engineering, which allows for large division factors. By optically coupling the OPO output to a modelocked soliton comb, synchronization can occur, which locks the soliton repetition rate to a fraction of the OPO mode spacing despite their disparate temporal waveforms (Fig. \ref{figScheme}E). Discrete narrow-linewidth microwave tones are generated with the lowest frequency component (non-DC) being the soliton repetition rate. In our proof-of-concept experiment, we use a signal-idler frequency separation of 8 THz and a 200-GHz soliton comb to achieve a 20-dB repetition-rate-noise reduction of the synchronized state compared to free-running solitons that have previously been used for microwave generation \cite{Liu_NP_20,Weng_SciAdv_20}.

A critical performance metric of a microwave source is the phase noise, which is associated with time-dependent deviations of the source frequency from the nominal carrier frequency. Such phase fluctuations lead to imperfections in timing and frequency and impose an upper limit on the performance for many applications. The phase noise of a microwave signal resulting from OFD is fundamentally limited by that of the dual-point optical reference, which needs to be generated with ultra-low noise. It has been shown that in a nonlinear microresonator temporal patterns that fully fill the cavity, such as the OPO and Turing rolls, exhibit higher phase stability than localized patterns, such as cavity solitons \cite{Coillet_OL_14}. This is mainly due to Kerr solitons existing in the red pump-cavity-detuning regime, which has higher nonlinear and thermal instabilities \cite{Drake_NatPhot_20}. In addition, higher-order dispersion and Raman nonlinearity can lead to increased soliton-timing jitter \cite{Matsko_JOSAB_15,Yi_NatComm_17,Yang_NatComm_21} and higher sensitivity to thermal noise \cite{Bao_OL_17}. In contrast, OPOs exist in the blue-detuned regime and do not suffer from dispersive-wave formation or self-frequency shifts induced by higher-order dispersion and Raman nonlinearity, which suggests a promising solution for a stable, on-chip dual-point optical reference. Furthermore, the higher optical powers of OPOs allow for a narrow Schawlow-Townes linewidth (STL), which represents the fundamental spectral linewidth due to quantum fluctuations. Lastly, energy conservation imposes strong frequency correlations between the signal and idler modes, which can be exploited for OFD. We theoretically and experimentally demonstrate three key noise-suppression features of OPO including low STL, strong pump-noise rejection, and strong thermal-noise suppression, which make it an ideal dual-point optical reference for OFD. Furthermore, the energy-conservation requirement of the OPO resembles that of the soliton, making it possible to achieve synchronization via simple photonic coupling. In a previous work, a electro-optical comb was electronically locked to a Turing roll in a MgF$_2$ resonator \cite{Weng_PRA_21}. However, for the pump-cavity detuning used in \cite{Weng_PRA_21}, the high stability similar to what OPO can possess was not observed, and the performance was on par with free-running solitons on similar platforms \cite{Yang_NatComm_21}.

The phase $\psi_m$ of the microwave signal generated by OFD can be expressed as,
\begin{equation}\label{eq_OFD}
	\psi_m = \frac{\psi_i-\psi_s}{N},
\end{equation}
where $N$ is the division factor, and $\psi_s$ and $\psi_i$ are the phases of the OPO signal and idler, respectively, which can be described as random processes resulting from phase diffusion. The resulting power spectrum of $\psi_m$ shows a strong noise reduction by a factor of $1/N^2$ due to OFD and a complete rejection of common-mode fluctuations between the signal and idler phases, which constitute a large part of the classical noise in OPOs. The fundamental phase-noise limit corresponds to the STL of the signal and idler, which can be reached with sufficient common-mode-noise rejection. Based on the classical model and assuming the signal and idler have identical resonator $Q$s, their phase difference is described by,
\begin{equation}\label{eq_Classical}
	\frac{d}{dt}(\psi_i-\psi_s) = (k_s-k_i)\Delta T,
\end{equation}
where $\Delta T$ is the time-dependent temperature variation, and $k_s$, $k_i$ are the coefficients of thermal-induced resonance shift for the signal and idler resonances, respectively. Notably, the pump noise is fully suppressed in the signal-idler phase difference. Such pump-noise suppression has been previously discussed in $\chi^{(2)}$- and $\chi^{(3)}$-based OPOs in the absence of thermal effects \cite{Courtois_JMO_91,Matsko_JOSAB_15}, and similar principles have been used for low-noise microelectromechanical oscillators \cite{Kenig_PRL_12}. Here, we show that this suppression can be leveraged for OFD-based low-noise microwave generation. In general, $k_s$ and $k_i$ are proportional to the signal and idler frequencies, and thus have a non-zero difference. However, many low-thermorefractive-waveguide designs can be implemented to further mitigate this effect \cite{Guha_OE_13,Djordjevic_OE_13,Rodrigues_CLEO_22}. We show theoretically that by incorporating a small amount of material with the opposite thermal coefficient of the core material, $k_s=k_i$ can be achieved even for far-separated frequencies (see Supplementary Material). Notably, the $k$ coefficient only needs to be matched at two wavelength points, which is achievable even for large wavelength separations. In addition, the OPO operates in the blue-detuned regime, which allows for laser cooling via thermo-optical backaction using sufficiently low-noise pump lasers \cite{Diosi_PRA_08,SafaviNaeini_NJP_13,Sun_PRA_17}. Thus, the OPO provides an efficient approach to suppress classical noise sources that are intrinsic to the pump laser and waveguide material.

We determine the STL of the OPO signal, idler, and dual-point reference by performing a fully quantum-optical analysis of triply-resonant $\chi^{(3)}$-based OPO (Supplementary Material). We show that vacuum fluctuations result in a Lorentzian lineshape similar to those of lasers and $\chi^{(2)}$ OPOs \cite{Yamamoto_PRA_90} with a full-width-at-half-maximum (FWHM) linewidth of the dual-point reference given by, 
\begin{equation}\label{eq_STL}
	\Delta f_\mathrm{ST} = \frac{\hbar\omega_s\kappa(\alpha^2+\sigma^2)}{2\alpha P_s},
\end{equation}
where $\hbar$ is Planck's constant, $\omega_s$ is the angular frequency of the signal, $P_s$ is the output power of the signal, $\kappa$ is the output coupling rate, $\alpha$ is the cavity loss rate, and $\sigma$ is an additional linewidth broadening term due to the nonlinear-phase shift and phase mismatch (Supplementary Material). The individual linewidths of the signal and idler are $\Delta f_\mathrm{ST}/4$ due to the phase correlations. Owing to the availability of high $Q$ microresonators and the high output power of OPO states, sub-hertz STLs are readily achievable. Moreover, the dual-point reference linewidth does not depend on its spectral separation, making it compatible with OFD with a large division factor. 

\begin{figure*}
	\centering
	\includegraphics{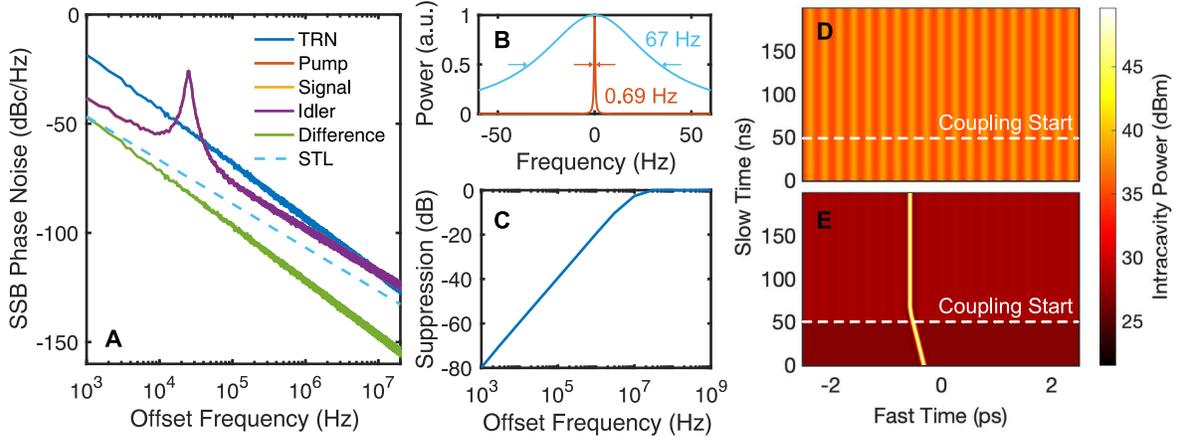}
	\caption{\textbf{Numerical simulation of OPO noise and OPO-soliton synchronization.} (A) Numerical simulation of an OPO generated in a 200-GHz-FSR microresonator with a loaded $Q$ of $10^6$. The pump and thermal noise are input parameters given by the red and blue traces, which correspond to our experimental measurements. The signal and idler phase noise are calculated via numerical simulation (yellow and purple). The green trace corresponds to the noise of the difference phase between the signal and idler modes, which shows strong suppression of the pump and thermal noise. The dashed line corresponds to the dual-point STL given by Eq. (\ref{eq_STL}). (B) The optical power spectrum of STL-limited fields. The blue trace corresponds to the STL trace in (A), whereas the red trace corresponds to the case of a 50-GHz-FSR microresonator with a $Q$ of $4\times10^6$, which yields a sub-hertz dual-point STL. (C) The maximum phase-noise reduction as a function of frequency for all-optical synchronization. The microresonator parameters are identical to those in (A). (D, E) Temporal evolution of the (D) OPO and (E) soliton comb under all-optical synchronization. The drift of the soliton timing is captured by the OPO trajectory after the optical coupling is turned on, with a coupling coefficient of $2.25\times10^{-6}$ per roundtrip time.}
	\label{figSimulation}
\end{figure*}

We numerically simulate the classical noise performance using stochastic equations, in which the thermorefractive noise (TRN) is modeled based on the experimental characterization of our SiN device (see Supplemental Material). We assume a Lorentzian-lineshape pump with an FWHM linewidth of 2 kHz and a noise sidelobe at 22-kHz offset frequency, which represents the condition in our experiment. We ignore the pump-induced thermorefractive backaction and simulate a microresonator with a free-spectral range (FSR) of 200 GHz and a loaded $Q$ of $10^6$ (see Supplementary Material). Figure \ref{figSimulation}A, shows the power spectral density of the single-side-band (SSB) phase noise for different components of OPO. The pump-noise sidelobe is eliminated in the relative-phase-noise spectrum carried by the beat note of the signal and idler fields, which corresponds to $>$47-dB noise suppression. The residual relative-phase noise follows the TRN spectrum with a 29-dB reduction, in agreement with the analytical result of Eq. (\ref{eq_Classical}). The quantum-noise limit of the dual-point reference is plotted in dashed lines corresponding to an STL of 67 Hz, which can be further reduced by increasing the $Q$ or the OPO power. For example, we can achieve a sub-Hertz dual-point STL by using a 50-GHz-FSR microresonator with a $Q$ of $4\times10^6$ and a pump power of 400 mW (Fig. \ref{figSimulation}B). 

We perform OFD by synchronizing a soliton modelocked Kerr comb to the stable reference provided by the OPO sidebands. Synchronization is an all-optical process that has been used to lock the comb spacing between two modelocked Kerr combs via optical coupling, which has been demonstrated for unidirectional coupling from one microresonator to another \cite{Jang_NatPhot_18,Jang_PRL_19,Kim_SciAdv_21}, and the comb in the coupled secondary microresonator inherits the phases of the comb in the primary microresonator. Previous demonstrations have focused on dynamically similar states of Kerr resonators, namely solitons \cite{Jang_NatPhot_18,Jang_PRL_19} and nonsolitonic combs \cite{Kim_SciAdv_21}. However, the OPO and soliton states are dynamically distinct, which can be shown using the bifurcation of the homogeneous (\textit{i.e.}, only keeping the pump mode) solutions plotted as a function of normalized intracavity power against normalized detuning \cite{Chembo_PRA_10,Coen_OL_13,Godey_PRA_14}. As shown in Fig. \ref{figScheme}B, the OPO state exists in the blue-detuned branch (left solid trace), while the soliton state exists in the red-detuned branch (right solid trace). The corresponding temporal waveforms are shown in Fig. \ref{figScheme}E, which have distinct shapes. The synchronization of such dynamically different waveforms has been an open question. 

We first simulate the OPO-soliton synchronization process using the model in \cite{Jang_NatPhot_18}. As shown in Fig. \ref{figSimulation}D and \ref{figSimulation}E, the deviation of the soliton-repetition rate from the desired value manifests as a drift in the fast-time frame. We numerically introduce a unidirectional power coupling of $7.3\times10^{-6}$ per roundtrip time (Supplementary Material), which traps the soliton peak to one of the OPO peaks. Consequently, the soliton-repetition rate is synchronized to a fraction of the OPO-mode spacing. We theoretically investigate the maximum noise suppression capability of this configuration by numerically implementing an absolutely stable OPO (Supplementary Material). Figure. \ref{figSimulation}C shows the ratio between the residue noise strength and the initial noise strength of the repetition rate of the soliton comb. The synchronization scheme provides limited noise suppression for noise at frequencies comparable to the cavity linewidth. However at lower offset frequencies, the noise suppression strength increases by 20 dB/decade, reaching -59 dB at 10-kHz offset frequency. Further improvement of the noise suppression bandwidth can be achieved by increasing the soliton-cavity linewidth or the coupling strength.

\begin{figure*}
	\centering
	\includegraphics{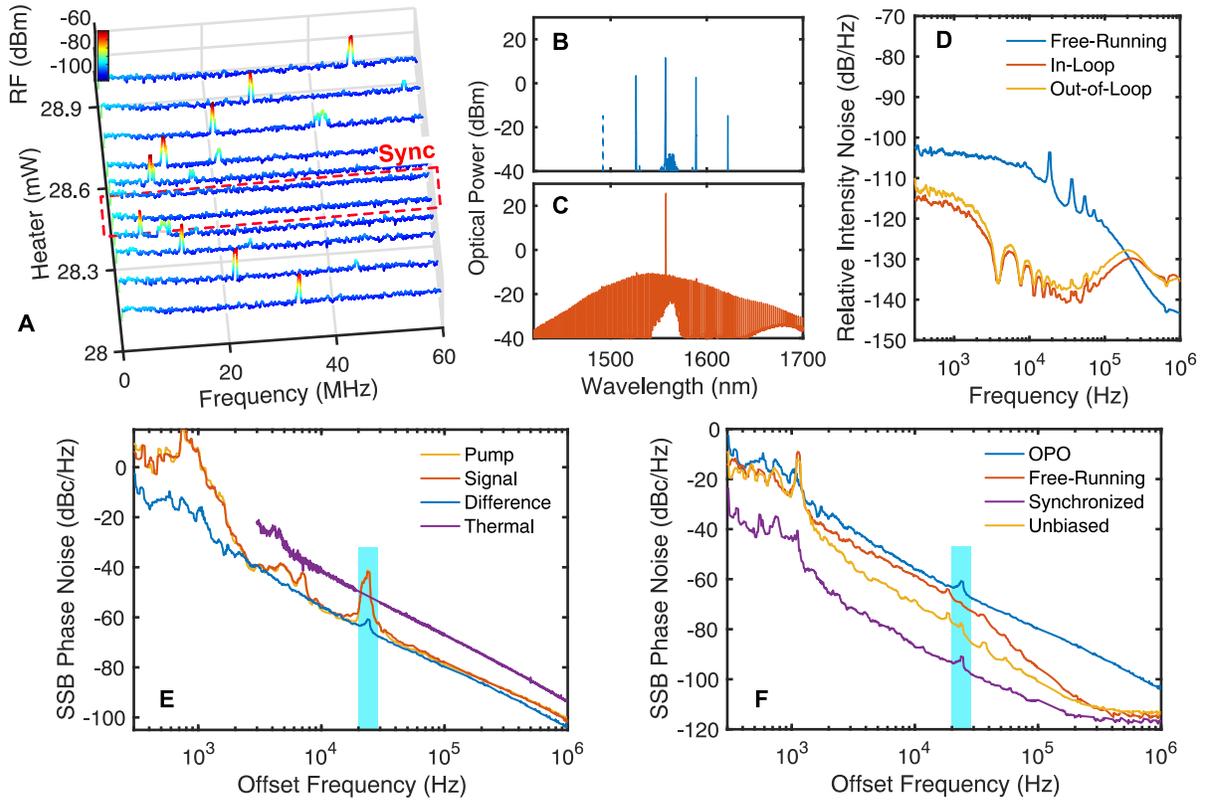}
	\caption{\textbf{Experimental demonstration of OPO-soliton synchronization}. (A) OPO-soliton beat note as the soliton heater is tuned. The synchronization regime is marked in dashed lines. (B, C) Optical spectra of the (C) OPO and (D) soliton comb, respectively. The bump around 1700 nm in (C) is an artifact from the spectrum analyzer. (D) The relative intensity noise of the pump laser before and after feedback control. (E) Noise characterization of the various components of the OPO, including the pump, signal, signal-idler difference frequency (7.7 THz), and the TRN noise. The strong noise peaks in the pump laser are suppressed in the relative-phase noise. (F) Comparison of the phase noise in the soliton repetition rate (227 GHz) when the soliton is free-running and synchronized to the OPO. For reference, the relative phase noise of the OPO is shown. The noise of a free-running soliton without temperature bias is shown in yellow, which has lower noise than the heated soliton but higher than the synchronized soliton.}
	\label{figSync}
\end{figure*}

In our experiment, we use a single CW source at 1557~nm to pump both the OPO and soliton-comb resonator. The two microresonators have an FSR of 227~GHz and are identical in design. The cross-section of all our SiN waveguides measures 730$\times$1500~$\mathrm{nm^2}$ with a ring-bus coupling gap of 350 nm and a gap for the coupling link being 450 nm, which provides a much lower coupling rate than the intrisic loss of the the cavity. A single pump laser is split on-chip with 24 mW going into the OPO ring and 370 mW going into the soliton ring. The soliton pump power can be reduced in future designs by adjusting the on-chip slitting ratio for more power efficient operations. The sidebands of the OPO are located at 1526.5~nm and 1588.8~nm, corresponding to a frequency separation of 7.7~THz and a mode separation of 34. The optical spectra of the OPO and soliton comb are shown in Fig. \ref{figSync}B and \ref{figSync}C. To facilitate synchronization, we monitor the beat note between the OPO and the soliton line at 1588.8 nm. This beat note can be tuned by tuning the heater power applied to the soliton (Fig. \ref{figSync}A), which is sensitive to thermal fluctuations. The existence of a beat note indicates the OPO and soliton running independently, which corresponds to a heater power $<$ 28.3 mW or $>$ 28.6 mW and a beat note frequency $>$ 11 MHz. Near the two threshold heater powers, the beat note jitters rapidly between DC and 11 MHz, indicating the soliton being captured by the OPO and then slipping away in short timescales. For a heater range between 28.4 mW and 28.5 mW, the beat note vanishes, indicating stable synchronization. Based on the heater values, we infer a capturing range of $\approx$ 10 MHz, which can be further increased by adopting stronger coupling or higher OPO power. 

\begin{figure*}
	\centering
	\includegraphics{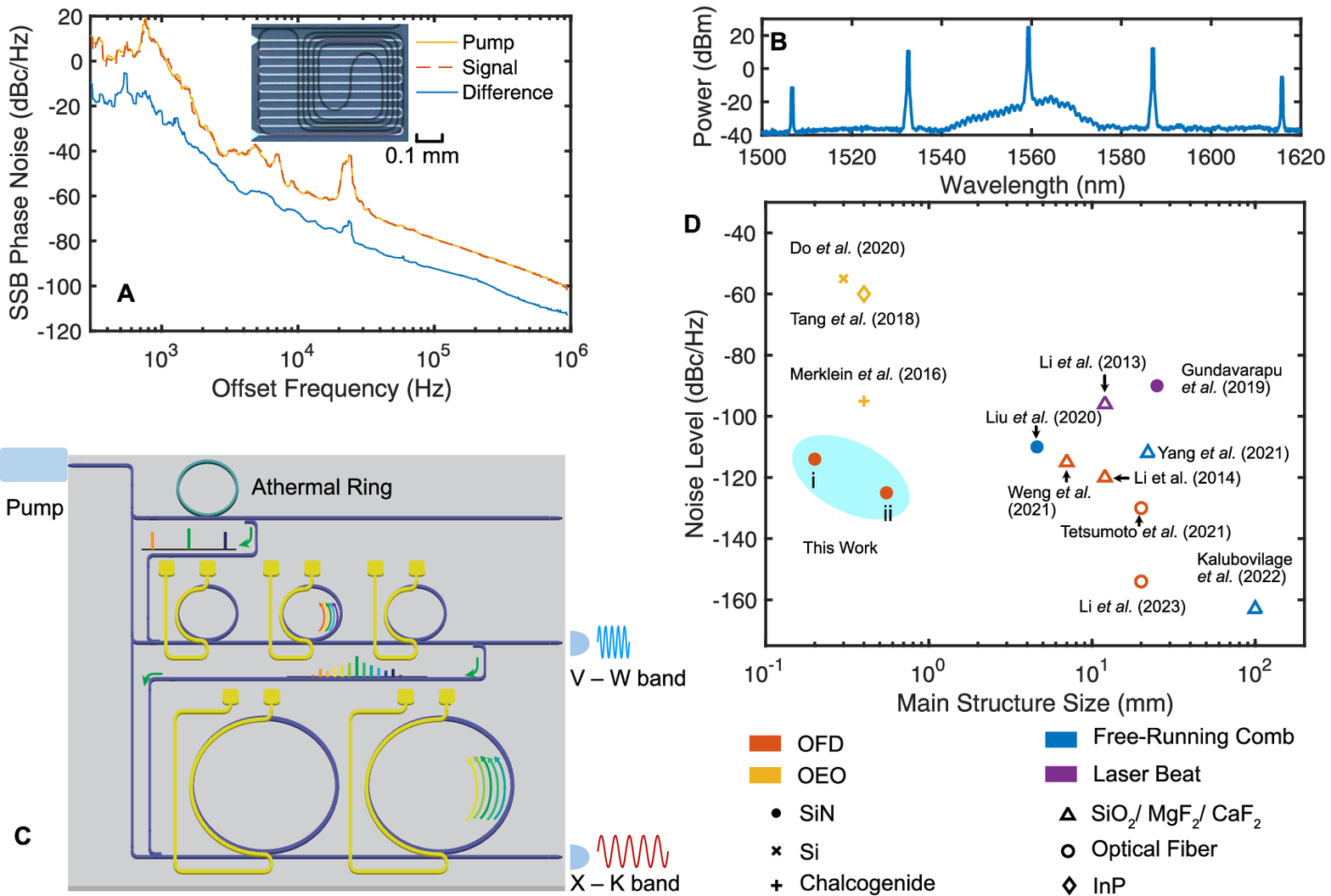}
	\caption{\textbf{Improved OPO performance and proposed full system for tunable microwave generation}. (A, B) The (A) optical spectrum and (B) phase noise of the OPO generated in a 20-GHz-FSR microresonator. We measure a noise level of -68 dBc/Hz at 10-kHz offset with a carrier frequency of 6.7 THz. The inset of (A) shows the microscope image of the OPO cavity. (C) Proposed on-chip ultra-low-phase-noise widely-tunable microwave source. The optical reference is a broad-band OPO generated in the athermal ring. Multiple soliton rings are fabricated to allow Kerr-comb generation with mode spacings in the microwave X to W band. On-chip heaters are used to activate the desired ring with optical detuning. The microwave signal is generated via photodetection of the Kerr-comb signals. (D) Comparison of the recent progresses of microwave generation using optical devices, plotted as phase noise at an equivalent 10-GHz carrier frequency and 10-kHz offset frequency against the size of the main structure \cite{Merklein_OL_16,Tang_OE_18,Do_SciRep_20,Li_NatComm_13,Li_Science_14,Gundavarapu_NatPhot_19,Tetsumoto_NatPhot_21,Li_Optica_23,Weng_PRA_21,Liu_NP_20,Yang_NatComm_21,Kalubovilage_OE_22}. The current work for (i) synchronized 227-GHz soliton and (ii) 6.7-THz OPO is highlighted in the blue-shaded region.}
	\label{figVision}
\end{figure*}

An important result of synchronization is reduced soliton-phase noise, which we measure using a modified delayed self-heterodyne technique that enables measurement of both the absolute phase noise of the individual optical fields and the relative phase noise between the two components (Fig. \ref{figScheme}D) \cite{Kwon_SciRep_17,Tian_OE_20}. Figure \ref{figSync}E shows the SSB phase noise of the OPO, where the yellow and red curves correspond to the phase noise of the pump and the OPO signal (1588.8~nm), respectively, and the blue curve corresponds to the relative phase noise between the OPO signal and idler. The pump noise has a strong peak at 22~kHz (blue-shaded region), which is reduced by 22 dB in the relative phase noise as predicted by our theoretical analysis [Eq. (\ref{eq_Classical})]. The residual noise is due to the slight cavity lifetime difference between the signal and idler modes due to wavelength-dependent ring-bus coupling. We also characterize the room-temperature TRN of our microresonator using the homodyne measurement proposed in \cite{Gorodetksy_OE_10}, which yields a result that largely agrees with the experimental characterization shown in \cite{Huang_PRA_19}. The OPO relative noise is found to be 13 dB lower than the TRN due to common-mode noise rejection. We attribute the current limit of OPO phase noise to the relatively large intensity and phase noise of our pump laser, which leads to optical heating instead of cooling \cite{Diosi_PRA_08,SafaviNaeini_NJP_13,Sun_PRA_17} via the thermorefractive effect. We also note that an off-chip laser intensity stabilization loop is employed in the current experiment using a general purpose proportional-integral-derivative (PID) controller (Fig. \ref{figSync}D) and a photodetector. A stronger noise suppression can be achieved using customized algorithms. Employing a microresonator with a larger mode volume or athermal designs (Supplementary Material) can also reduce the effect of TRN and intensity noise \cite{Huang_PRA_19,Guha_OE_13,Djordjevic_OE_13,Rodrigues_CLEO_22}. Figure \ref{figSync}F shows the relative phase noise between two adjacent soliton-comb lines when the soliton is free-running (red) and is synchronized to the OPO (purple). In the latter case, we observe a 28-dB reduction in the phase noise, confirming the large phase-noise reduction with OPO-soliton synchronization. Furthermore, we do not observe a strong noise recoil that presents in electronic locking systems where, at certain frequency bands, the negative feedback turns into positive feedback. For reference, we also plot the relative phase noise of the OPO sidebands in blue, which is 31 dB higher than the relative noise between adjacent soliton lines, in agreement with the division factor $N^2=1156$. The noise-reduction bandwidth exceeds 300 kHz, which is only limited by the detector noise floor rather than the synchronization process. The free-running soliton noise is limited by the control signal applied to the integrated heater (see Supplementary Material) which is necessary for soliton generation \cite{Joshi_OL_16}. The OPO is generated at low heater voltage due to the more robust generation process. However, a small TRN increase is introduced to the OPO ring due to the thermal cross talk from the soliton ring. We also characterize the soliton noise without applied heat by first generating the soliton using thermal tuning, then jointly tuning the laser wavelength and the heater until the heater can be turned off without destroying the soliton state. This is shown as the yellow trace in Fig. \ref{figSync}F, where the noise is nonetheless 18 dB higher than the synchronized soliton.

Microwave phase noise is often characterized at the 10-kHz offset frequency by scaling the carrier frequency to 10 GHz. The free-running soliton in this experiment corresponds to a noise level of -86 dBc/Hz, whereas the synchronized soliton corresponds to -114 dBc/Hz. The latter is 4 dB lower than the 10-GHz microwave generated by the free-running soliton in \cite{Liu_NP_20}, which was the lowest based on the SiN platform. Notably, the device used in this experiment has an inherently larger TRN than \cite{Liu_NP_20} due to a significantly smaller mode volume. We also characterize the OPO noise of a larger-volume resonator (FSR = 20 GHz), which is shown in Fig. \ref{figVision}A and \ref{figVision}B. We infer a 10-GHz equivalent noise of -125 dBc/Hz, which is 15-dB lower than \cite{Liu_NP_20} and also lower than OFD with chip-based Brillouin lasers \cite{Li_Science_14}. It is also worth noting that our 20-GHz cavity occupies an area of 566 $\times$ 417 $\mu$m$^2$ (Fig. \ref{figVision}A, inset), which is much smaller than typical Brillouin laser cavities. We also examine the transfer of phase noise from the pump to the soliton, which is the dominant noise source for free-running 10-GHz solitons \cite{Liu_NP_20}, where a transfer coefficient of -55 dB was observed. Our 227-GHz-soliton noise is largely limited by TRN. However, using the pump noise peak at 22 kHz (blue-shaded region), we infer a pump-to-soliton-noise transfer coefficient of -39 dB for the free-running soliton without heater bias. As a comparison, the transfer coefficient is -51 dB for the synchronized soliton at 227 GHz, and we infer a transfer coefficient of -78 dB if a 10-GHz soliton is synchronized based on the N$^2$ noise scaling. Further reduction of noise can be achieved with athermal designs, where a thermorefractive coefficient reduction of more than 100$\times$ has been demonstrated \cite{Raghunathan_OE_10}, corresponding to a TRN reduction of 40 dB. The STL limit can be reduced by increasing the power of OPO or the signal-idler separation via dispersion engineering, which increases the division factor. 10-GHz microwave can reach a noise level around -170 dBc/Hz using bulk optical equipments such as monolithic-cavity modelocked lasers \cite{Kalubovilage_OE_20,Kalubovilage_OE_22} or fully stabilized combs \cite{Xie_NatPhot_17}. Comparable performance (<-165 dBc/Hz) can be envisioned with our on-chip-OFD scheme with a TRN reduction of 40 dB, OPO span of 50 THz, and dual-point STL of 0.2 Hz, which are readily achievable with the current technological capabilities. Figure \ref{figVision}D shows a comparison of the recent advances of microwave generation in compact optical devices, plotted as the phase noise at the 10-GHz or equivalent carrier frequency and 10-kHz offset frequency versus the size of the main structure that provides the phase stability, including the optical delay for optoelectronic oscillators (OEO) \cite{Merklein_OL_16,Tang_OE_18,Do_SciRep_20}, Brillioun laser cavites \cite{Li_NatComm_13,Li_Science_14,Gundavarapu_NatPhot_19,Tetsumoto_NatPhot_21,Li_Optica_23}, OPO (this work) or Turing roll \cite{Weng_PRA_21} cavites, and frequency-comb cavities for the free-running scheme \cite{Liu_NP_20,Yang_NatComm_21,Kalubovilage_OE_22}. The pump lasers and electro-optical modulators are not included. Optical fibers are assumed to occupy an area of 2 cm in diameter which results in a loss of $<$ 0.5 dB/loop for the typical SMF-28\textsuperscript{\textregistered} fiber. As shown in the figure, larger structural size yields better noise performance as the TRN is reduced. However, our synchronized 227-GHz soliton (dot i) achieves the lowest phase noise on the SiN platform while occupying the smallest footprint. A further 11-dB improvement can be achieved by increasing the OPO footprint (dot ii), which surpasses the performance of lower-TRN platforms such as SiO$_2$ and MgF$_2$-based whispering-gallery-mode resonators.

Based on the synchronization of low-noise OPO and Kerr comb, a compact, ultra-low-noise, broadly tunable, high-frequency microwave oscillator can be envisioned. Figure \ref{figVision}C shows a possible design of a full system. A broad-band reference OPO is generated in the top athermal ring which is evanescently coupled to the middle bus waveguide to facilitate synchronization. Multiple microresonators with different mode spacings in the microwave V to W band are used for broad-band soliton generation. The on-chip heaters are used to control which microresonator is activated via frequency detuning. V to W band microwave frequencies are generated upon photodetection of the middle-level comb. To further divide the frequency to the microwave X to K bands, we can implement low-FSR microresonators on the bottom level, which can have a narrow bandwidth to achieve higher efficiency \cite{Jang_OL_21,Kondratiev_arXiv_22}. However, narrow bandwidth combs result in small division factors which can limit the achievable phase noise. This can be addressed by synchronizing the low-FSR comb to the larger-FSR comb, which corresponds to a harmonic synchronization scheme \cite{Jang_PRL_19}. The cascading allows the narrow-band low-FSR comb to have an effective bandwidth identical to the larger-FSR comb to achieve a large division factor and hence low phase noise after OFD. X to K band microwave signals are generated upon photodetection of the bottom-level comb.

In conclusion, we have demonstrated an ultra-compact frequency division scheme based on passive synchronization of a low-noise OPO and Kerr-soliton comb, which requires only a single pump laser. Our demonstration shows, for the first time, that dynamically different states in Kerr cavities can be passively synchronized via physical coupling. In particular, a phase-noise reduction of 28 dB is observed between free-running and phase-referenced solitons, which enabled us to reach a record-low soliton-phase noise on the SiN platform. In addition, we have shown that the OPO state can serve as ultra-stable optical references due to three key features, namely low STL, strong pump-noise suppression, and strong thermal-noise suppression. Our OFD scheme allows for high performance microwave generation using a single laser with modest noise performance, which provides a promising pathway toward small-footprint ultra-low-noise microwave generation.

\textbf{Acknowledgements:}  This work was performed in part at the Cornell Nano-Scale Facility, which is a member of the National Nanotechnology Infrastructure Network, supported by the NSF, and at the CUNY Advanced Science Research Center NanoFabrication Facility. We acknowledge computing resources from Columbia University's Shared Research Computing Facility project, which is supported by NIH Research Facility Improvement Grant 1G20RR030893-01, and associated funds from the New York State Empire State Development, Division of Science Technology and Innovation (NYSTAR) Contract C090171, both awarded April 15, 2010. 

\textbf{Funding:} This work was supported by Defense Advanced Research Projects Agency of the U.S. Department of Defense (Grant No. HR0011-22-2-0007 ), Army Research Office (ARO) (Grant No. W911NF-21-1-0286), and Air Force Office of Scientific Research (AFOSR) (Grant No. FA9550-20-1-0297). 

\textbf{Author contributions:} Y.Z., Y.O., and A.L.G conceived the project. Y.Z. and J.K.J. performed the theoretical analysis and experiment. Y.Z., J.K.J., Y.O., and A.L.G. performed the data analysis with input from all authors. X.J. and K.J.M. fabricated the silicon-nitride devices under the supervision of M.L. Y.Z., J.K.J. and A.L.G. wrote the manuscript with feedback from all authors. 

\textbf{Competing interests:} The authors declare no competing interests. 

\textbf{Data and materials availability:} All data is available in the manuscript or the supplementary materials. Simulation code may be obtained from the authors upon reasonable request

\section{Methods}
\textbf{Phase-Noise Characterization System}
The phase-noise characterization system is shown in Fig. 1D, which consists of an imbalanced Mach-Zehnder interferometer (MZI) and an electronic-mixing stage. A fiber delay of 200 m is used to extend the length of the bottom MZI arm, and an acusto-optic oscillator (AOM) is placed in the top arm of the MZI to shift the optical frequency by 80 MHz. A fiber-based polarization controllers is used to match the polarization of the two MZI arms, which is not shown in the figure. The fields at the end of the upper and lower MZI arms can be written as,
\begin{align}
	&\tilde{E}_u = Ee^{i(\omega+\Delta\omega)t + i\tilde{\psi}(t)+i\tilde{\phi}_\mathrm{AOM}(t)},\\
	&\tilde{E}_l = Ee^{i\omega(t+\tau)+i\tilde{\psi}(t+\tau)+i\tilde{\phi}_\mathrm{fiber}(t)},
\end{align}
where $\omega$ is the laser carrier frequency, $\Delta\omega$ is the 80-MHz AOM frequency shift, $\tau$ is the time delay through the 200-m fiber delay, $\tilde{\psi}(t)$ is the laser phase at time $t$ excluding the carrier phase, $\tilde{\phi}_\mathrm{AOM}(t)$ is the AOM phase at time $t$ excluding the carrier phase, and $\tilde{\phi}_\mathrm{fiber}(t)$ is the added phase due to fiber fluctuations. The detected AC-coupled voltage after the fiber beam splitter is,
\begin{align}
	&V \propto \notag\\&\cos\left[\Delta\omega t +\omega\tau+\tilde{\psi}(t)-\tilde{\psi}(t+\tau)+\tilde{\phi}_\mathrm{AOM}(t)-\tilde{\phi}_\mathrm{fiber}(t)\right].
\end{align}
where $\omega\tau$ is a fixed phase offset that does not affect the noise. The phase noise of $V$ can be expressed as,
\begin{align}
	\mathcal{L}_{V}(f) = 4\sin^2(\pi\tau f)\mathcal{L}_{\psi}(f) + \mathcal{L}_\mathrm{AOM}(f) + \mathcal{L}_\mathrm{fiber}(f),
\end{align}
where $\mathcal{L}_{\psi}$, $\mathcal{L}_\mathrm{AOM}$ and $\mathcal{L}_\mathrm{fiber}$ are the phase noise spectra of the laser, AOM, and fiber, respectively. Here, we have assumed that the laser, AOM, and fiber noise are uncorrelated due to their different origins. In general, the AOM noise is much lower than the laser noise due to its lower carrier frequency, which can be ignored. The fiber noise is high at low offset frequencies but quickly reduces for frequencies above 1 kHz. Thus, $\mathcal{L}_{V}$ reflects the laser noise up to a known sinusoidal modulation with increased sensitivity for higher offset frequencies.

To measure the noise correlation of two wavelengths, we first measure their self-heterodyne beatnote with separate photodetectors to get,
\begin{align}
	V_1 \propto \cos\left[\Delta\omega t+\tilde{\psi}_1(t)-\tilde{\psi}_1(t+\tau)+\tilde{\phi}_\mathrm{AOM}(t)-\tilde{\phi}_\mathrm{fiber}(t)\right],
\end{align}
and,
\begin{align}
	V_2 \propto \cos\left[\Delta\omega t+\tilde{\psi}_2(t)-\tilde{\psi}_2(t+\tau)+\tilde{\phi}_\mathrm{AOM}(t)-\tilde{\phi}_\mathrm{fiber}(t)\right],
\end{align}
where we have ignored the constant phase terms. We then upshift the frequency of $V_1$ using a frequency mixer and a local oscillator (LO) at 109 MHz, which yields,
\begin{align}
	V_1' \propto &\cos\left[(\Delta\omega+\omega_\mathrm{LO}) t+\tilde{\psi}_1(t)-\tilde{\psi}_1(t+\tau)\right.\notag\\
	&\left.+\tilde{\phi}_\mathrm{AOM}(t)-\tilde{\phi}_\mathrm{fiber}(t)+\tilde{\phi}_\mathrm{LO}(t)\right].
\end{align}
A second frequency mixer performs difference-frequency generation between $V_1'$ and $V_2$, which yields,
\begin{align}
	V_3 \propto &\cos\left[\omega_\mathrm{LO} t+\tilde{\psi}_1(t)-\tilde{\psi}_1(t+\tau)-\tilde{\psi}_2(t)\right.\notag\\
	&\left.+\tilde{\psi}_2(t+\tau)+\tilde{\phi}_\mathrm{LO}(t)\right],
\end{align}
which has removed the AOM and fiber noise. The phase noise of $V_3$ corresponds to,
\begin{align}
	\mathcal{L}_{V3}(f) = 4\sin^2(\pi\tau f)\mathcal{L}_{\Delta\psi}(f) + \mathcal{L}_\mathrm{LO}(f),
\end{align}
where $\mathcal{L}_{\Delta\psi}$ is the relative phase noise between the two lasers, and $\mathcal{L}_\mathrm{LO}(f)$ is the phase noise of the LO. With a sufficiently low LO noise, $\mathcal{L}_{V3}$ gives a direct measurement of the relative phase noise of the lasers up to a known sinusoidal modulation.	


	\newpage
	\onecolumngrid
	\setcounter{equation}{0}
	\setcounter{figure}{0}
	\setcounter{table}{0}
	\setcounter{page}{1}
	\setcounter{section}{0}
	\renewcommand{\theequation}{S\arabic{equation}}
	\renewcommand{\thefigure}{S\arabic{figure}}
	\renewcommand{\bibnumfmt}[1]{[S#1]}
	\renewcommand{\citenumfont}[1]{S#1}

\begin{center}
\textbf{\large All-optical frequency division on-chip using a single laser: supplementary material}
\end{center}
\section{Schawlow-Townes Linewidth of $\chi^{(3)}$-Based OPO}\label{sec_STL}
The Schawlow-Townes linewidth (STL) represents the fundamental linewidth of an optical oscillator, which is induced by the quantum fluctuations of related fields. Due to the extremely low occupation number of thermal photons at room temperatures, the only quantum fluctuations contributing to the STL of an OPO are vacuum fluctuations. The $\chi^{(3)}$-based OPO process with vacuum fluctuations can be modeled as,
\begin{align}
	&\frac{d\hat{A}}{dt} = -\frac{\alpha}{2}\hat{A} - i\Delta_A\hat{A} + i\Gamma(\hat{A}^\dagger\hat{A}+2\hat{B}^\dagger\hat{B}+2\hat{C}^\dagger\hat{C})\hat{A} + i2\Gamma \hat{A}^\dagger\hat{B}\hat{C} + \sqrt{\kappa}A_\mathrm{in}+\sqrt{\alpha}\hat{a}_\mathrm{in},\label{eq_OPO1}\\
	&\frac{d\hat{B}}{dt} = -\frac{\alpha}{2}\hat{B} - i\Delta_B\hat{B} + i\Gamma(2\hat{A}^\dagger\hat{A}+\hat{B}^\dagger\hat{B}+2\hat{C}^\dagger\hat{C})\hat{B} + i\Gamma \hat{C}^\dagger\hat{A}^2 + \sqrt{\alpha}\hat{b}_\mathrm{in},\label{eq_OPO2}\\
	&\frac{d\hat{C}}{dt} = -\frac{\alpha}{2}\hat{C} - i\Delta_C\hat{C} +i\Gamma(2\hat{A}^\dagger\hat{A}+2\hat{B}^\dagger\hat{B}+\hat{C}^\dagger\hat{C})\hat{C}+i\Gamma \hat{B}^\dagger\hat{A}^2 + \sqrt{\alpha}\hat{c}_\mathrm{in},\label{eq_OPO3}
\end{align}
where $\hat{A}$, $\hat{B}$, and $\hat{C}$ are the annihilation operators of the cavity modes for the pump, signal, and idler fields, respectively, $\alpha$ is the loss rate of the cavity, $\Delta_A$, $\Delta_B$, and $\Delta_C$ are the detunings of the pump, signal and idler modes, respectively, $\Gamma$ is the cavity-enhanced nonlinear coefficient, $\kappa$ is the input coupling rate, $A_\mathrm{in}$ is the input field as a c-number, and $\hat{a}_\mathrm{in}$, $\hat{b}_\mathrm{in}$, and $\hat{c}_\mathrm{in}$ are the Langevin noise operators resulting from coupling to reservoirs with continuous mode spectra. We defined the detunings such that $\Delta_{A,B,C}>0$ when the field is red detuned. The nonlinear coefficient can be expressed as,
\begin{align}
	\Gamma =  \frac{3\hbar\omega_A^2\chi^{(3)}}{4\varepsilon_0n_\mathrm{eff}^2n_g^2V},
\end{align}
where $\hbar$ is Planck's constant, $\omega_A$ is the pump angular frequency, $\chi^{(3)}$ is the Kerr nonlinear coefficient, $\varepsilon_0$ is the vacuum dielectric coefficient, $n_\mathrm{eff}$ is the effective index of the pump, $n_g$ is the group index of the pump, and $V$ is the mode volume. Since the vacuum fluctuations are much weaker than the mean photon numbers in the OPO state, we can linearize Eqs. (\ref{eq_OPO1}) - (\ref{eq_OPO3}) by rewriting the operators in terms of their mean values and fluctuations as,
\begin{align}
	\hat{O} = (O + \hat{u}_o+i\hat{v}_o)e^{i\phi_o},
\end{align}
where $O\in\{A, B, C\}$. We choose the convention such that $O$ is a positive number and $\hat{u}_o$ and $\hat{v}_o$ are Hermitian operators. Similarly, we expand $\hat{a}_\mathrm{in}$, $\hat{b}_\mathrm{in}$, and $\hat{c}_\mathrm{in}$ as,
\begin{align}
	\hat{o}_\mathrm{in} = \hat{g}_o + i\hat{h}_o,
\end{align}
where $o \in \{a, b,c\}$, and $\hat{g}_o$ and $\hat{h}_o$ are Hermitian operators satisfying the  relations,
\begin{align}
	&\braket{\hat{g}_o(t)\hat{g}_o(t')} = \braket{\hat{h}_o(t)\hat{h}_o(t')} = \frac{1}{4}\delta(t-t'),\\
	&\braket{\hat{g}_o(t)\hat{h}_o(t')} = \braket{\hat{h}_o(t')\hat{g}_o(t)}^\ast = \frac{i}{4}\delta(t-t').
\end{align}
The operator equations after linearization reads,
\begin{align}
	\frac{d}{dt}\begin{pmatrix}
		\hat{u}_a \\ \hat{v}_a \\ \hat{u}_b \\ \hat{v}_b \\ \hat{u}_c \\ \hat{v}_c
	\end{pmatrix} = \mathbf{M}\begin{pmatrix}
	\hat{u}_a \\ \hat{v}_a \\ \hat{u}_b \\ \hat{v}_b \\ \hat{u}_c \\ \hat{v}_c
\end{pmatrix} + \sqrt{\alpha}\begin{pmatrix}
	\hat{g}_a \\ \hat{h}_a \\ \hat{g}_b \\ \hat{h}_b \\ \hat{g}_c \\ \hat{h}_c
\end{pmatrix},\label{eq_OPO_Lin}
\end{align}
where $\mathbf{M} = \{m_{ij}\}_{6\times 6}$ with coefficients $m_{11} = -\frac{\alpha}{2}+2\Gamma BC \sin\phi$, $m_{12} = -\Gamma(A^2+2B^2+2C^2+2BC\cos\phi)+\Delta_A$, $m_{13} = 2\Gamma AC\sin\phi$, $m_{14} = -2\Gamma AC\cos\phi$, $m_{15} = 2\Gamma AB\sin\phi$, $m_{16} = -2\Gamma AB\cos\phi$, $m_{21} = \Gamma(3A^2+2B^2+2C^2+2BC\cos\phi)-\Delta_A$, $m_{22} = -\frac{\alpha}{2}-2\Gamma BC\sin\phi$, $m_{23} = \Gamma(4AB+2AC\cos\phi)$, $m_{24} = 2\Gamma AC\sin\phi$, $m_{25} = \Gamma(4AC+2AB\cos\phi)$, $m_{26} = 2\Gamma AB\sin\phi$, $m_{31} = -2\Gamma AC\sin\phi$, $m_{32} = -2\Gamma AC\cos\phi$, $m_{33} = -\frac{\alpha}{2}$, $m_{34} = -\Gamma(2A^2+B^2+2C^2)+\Delta_B$, $m_{35} = -\Gamma A^2\sin\phi$, $m_{36} = \Gamma A^2\cos\phi$, $m_{41} = \Gamma(4AB+2AC\cos\phi)$, $m_{42} = -2\Gamma AC\sin\phi$, $m_{43} = \Gamma(2A^2+3B^2+2C^2)-\Delta_B$, $m_{44} = -\frac{\alpha}{2}$, $m_{45} = \Gamma(4BC+A^2\cos\phi)$, $m_{46} = \Gamma A^2\sin\phi$, $m_{51} = -2\Gamma AB\sin\phi$, $m_{52} = -2\Gamma AB\cos\phi$, $m_{53} = -\Gamma A^2\sin\phi$, $m_{54} = \Gamma A^2\cos\phi$, $m_{55} = -\frac{\alpha}{2}$, $m_{56} = -\Gamma(2A^2+2B^2+C^2)+\Delta_C$, $m_{61} = \Gamma(4AC+2AB\cos\phi)$, $m_{62} = -2\Gamma AB\sin\phi$, $m_{63} = \Gamma(4BC+A^2\cos\phi)$, $m_{64} = \Gamma A^2\sin\phi$, $m_{65} = \Gamma(2A^2+2B^2+3C^2)-\Delta_C$, and $m_{66} = -\frac{\alpha}{2}$, where $\phi = 2\phi_A-\phi_B-\phi_C$.

Equation (\ref{eq_OPO_Lin}) can be readily solved in the frequency domain with the Fourier transform,
\begin{align}
	\hat{o}(\omega) = \frac{1}{\sqrt{2\pi}}\int_{-\infty}^\infty \hat{o}(t)e^{i\omega t}dt,
\end{align}
where $o \in \{u_a, v_a, u_b, v_b, u_c, v_c, g_a, h_a, g_b, h_b, g_c, h_c\}$. The power spectral density of phase fluctuations of the signal can be evaluated as,
\begin{align}
	S(\omega) = \int \frac{\braket{\hat{v}_b(\omega)\hat{v}_b(\omega')}}{B^2} d\omega',
\end{align}
where $\hat{v}_b$ is a function of the frequency-domain noise operators satisfying,
\begin{align}
	&\braket{\hat{g}_o(\omega)\hat{g}_o(\omega')} = \braket{\hat{h}_o(\omega)\hat{h}_o(\omega')} = \frac{1}{4}\delta(\omega-\omega'),\\
	&\braket{\hat{g}_o(\omega)\hat{h}_o(\omega')} = \braket{\hat{h}_o(\omega')\hat{g}_o(
		\omega)}^\ast = \frac{i}{4}\delta(\omega-\omega').
\end{align}
In general, inverting $\mathbf{M}$ yields a complicated expression that needs to be evaluated numerically. However, in the current case where the signal and idler modes have identical loss rates, $\mathbf{M}$ can be simplified to yield a simple analytical approximation. First, we investigate the steady-state equations,
\begin{align}
	&\frac{\alpha}{2}A+i\Delta_AA-i\Gamma(A^2+2B^2+2C^2)A-i2\Gamma ABCe^{-i\phi} = \sqrt{\kappa}A_\mathrm{in}e^{-i\phi_A},\label{eq_OPO_Mean1}\\
	&\frac{\alpha}{2}B+i\Delta_BB-i\Gamma(2A^2+B^2+2C^2)B-i\Gamma A^2Ce^{i\phi} = 0,\label{eq_OPO_Mean2}\\
	&\frac{\alpha}{2}C+i\Delta_CC-i\Gamma(2A^2+2B^2+C^2)C-i\Gamma A^2Be^{i\phi} = 0.\label{eq_OPO_Mean3}
\end{align}
With some manipulation, we can find the following relations,
\begin{align}
	&B = C,\label{eq_OPO_Mean_Property1}\\
	&\Delta_B = \Delta_C,\label{eq_OPO_Mean_Property2}\\
	&\Gamma A^2\sin\phi = -\frac{\alpha}{2}.\label{eq_OPO_Mean_Property3}\\
	&\Delta_B-\Gamma(2A^2+3B^2) = \Gamma A^2\cos\phi.\label{eq_OPO_Mean_Property4}
\end{align}
Furthermore, the typical STL of an OPO is much lower than the cavity linewidth. Using the relation between phase-noise spectrum and laser linewidth \cite{DiDomenico_AO_2010}, we find that only the lowest order of $\omega$ contributes to the linewidth. With some calculation, we can find these terms as,
\begin{align}
	\bar{v}_b = i\frac{\alpha[\hat{h}_c(\omega)-\hat{h}_b(\omega)]-2(\Delta_B-2\Gamma A^2-2\Gamma B^2)[\hat{g}_c(\omega)-\hat{g}_b(\omega)]}{2\sqrt{\alpha}\omega}.
\end{align}
Subsequently, we can find the phase-noise spectrum as,
\begin{align}
	S(\omega) = \frac{\alpha^2+4(\Delta_B-2\Gamma A^2-2\Gamma B^2)^2}{4\alpha\omega^2B^2},
\end{align}
which correspond to the single-sideband (SSB) phase noise of,
\begin{align}
	\mathcal{L}(f) = \frac{\alpha^2+4(\Delta_B-2\Gamma A^2-2\Gamma B^2)^2}{8\pi\alpha f^2B^2},\label{eq_STL_SSB}
\end{align}
where $f$ is the offset frequency. This represents a Brownian diffusion process that is found in many oscillator systems. The corresponding linewidth can be found as,
\begin{align}
	\Delta f_\mathrm{ST} = \frac{\hbar\omega_B\kappa\left[\alpha^2+4(\Delta_B-2\Gamma A^2-2\Gamma B^2)^2\right]}{8\alpha P_B},\label{eq_STL}
\end{align}
where $P_B$ is the output power of the signal with $P_B = \hbar\omega_B\kappa B^2$.

\begin{figure}
	\centering
	\includegraphics{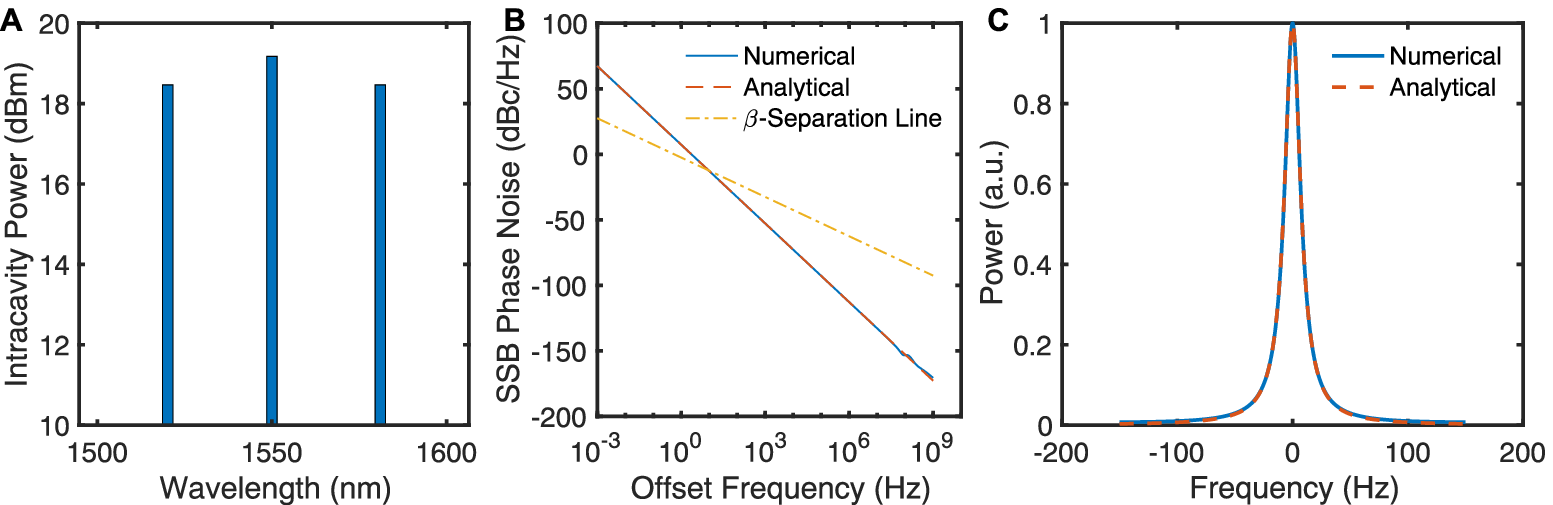}
	\caption{Theory of Schawlow-Townes linewidth of $\chi^{(3)}$ OPO. (A) Intracavity optical spectrum in the 3-mode model. (B) Comparison of vacuum-fluctuation-limited phase noise between the numerical solution of Eq. (\ref{eq_OPO_Lin}) (blue trace) and the analytical solution of (\ref{eq_STL_SSB}) (red trace). The $\beta$-separation line is also plotted, which indicates the part of phase noise that determines the laser linewidth. (C) The laser linewidth calculated by numerically integrating the blue trace in (B). The Lorentzian lineshape with an FWHM given in Eq. (\ref{eq_STL}) is shown as the red trace.}
	\label{figOPOTheroy}
\end{figure}

As an example, we simulate an OPO with $\alpha = 2\pi\times 200$ MHz, $\Delta_A = 2\pi\times 100$ MHz, and $\kappa = 2\pi\times 100$ MHz. We use 20 mW of pump power, corresponding to $|A_\mathrm{in}|^2 = 1.6\times10^{15}$ photons/s. In addition, we choose a free-spectral range of 200 GHz and a group-velocity dispersion (GVD) of -25 ps$^2$/km, corresponding to the OPO signal and idler at 1581.0 nm and 1520.2 nm, respectively, and $\Delta_B+\Delta_C-2\Delta_A = 1.97$ GHz. The OPO optical spectrum is shown in Fig. \ref{figOPOTheroy}A, where we have ignored the cascaded OPO lines due to their low powers. We numerically solve the phase-noise power spectrum using Eq. (\ref{eq_OPO_Lin}), which is shown as the blue trace in Fig. \ref{figOPOTheroy}B. The analytical solution is plotted as the red trace, which only slightly deviates from the numerical solution at offset frequencies comparable to the cavity linewidth. The $\beta$-separation line has been used to identify the phase noise that contributes to laser linewidths \cite{DiDomenico_AO_2010}, which is plotted in orange in Fig. \ref{figOPOTheroy}B. The laser linewidth is largely determined by the offset frequencies lower than the intersection point between the $\beta$-separation line and the noise spectrum, which is much lower than the cavity linewidth. This justifies our choice of keeping the lowest order of $\omega$. The STL-limited OPO lineshape can be calculated numerically by integrating the blue trace in Fig. \ref{figOPOTheroy}B \cite{DiDomenico_AO_2010}, which is shown as the blue trace in Fig. \ref{figOPOTheroy}C. The numerical result shows excellent agreement with a Lorentzian curve with full-width-at-half-maximum (FWHM) linewidth given by Eq. (\ref{eq_STL}).

\section{Classical Noise Sources of $\chi^{(3)}$-Based OPO}
The classical dynamics can be modeled analogously to the quantum dynamics of Eqs. (\ref{eq_OPO1})-(\ref{eq_OPO3}). We use $\sim$ to indicate classical random processes, which yields the dynamical equations,
\begin{align}
	&\frac{d\tilde{A}}{dt} = -\frac{\alpha}{2}\tilde{A} - i\tilde{\Delta}_A\tilde{A} + i\Gamma(\tilde{A}^\ast\tilde{A}+2\tilde{B}^\ast\tilde{B}+2\tilde{C}^\ast\tilde{C})\tilde{A} + i2\Gamma \tilde{A}^\ast\tilde{B}\tilde{C} + \sqrt{\kappa}\tilde{A}_\mathrm{in},\label{eq_OPO_Classical1}\\
	&\frac{d\tilde{B}}{dt} = -\frac{\alpha}{2}\tilde{B} - i\tilde{\Delta}_B\tilde{B} + i\Gamma(2\tilde{A}^\ast\tilde{A}+\tilde{B}^\ast\tilde{B}+2\tilde{C}^\ast\tilde{C})\tilde{B} + i\Gamma \tilde{C}^\ast\tilde{A}^2,\label{eq_OPO_Classical2}\\
	&\frac{d\tilde{C}}{dt} = -\frac{\alpha}{2}\tilde{C} - i\tilde{\Delta}_C\tilde{C} +i\Gamma(2\tilde{A}^\ast\tilde{A}+2\tilde{B}^\ast\tilde{B}+\tilde{C}^\ast\tilde{C})\tilde{C}+i\Gamma \tilde{B}^\ast\tilde{A}^2,\label{eq_OPO_Classical3}
\end{align}
where $\tilde{A}$, $\tilde{B}$, and $\tilde{C}$ are the field amplitudes normalized to photon number in the cavity, $\tilde{\Delta}_A$, $\tilde{\Delta}_B$, and $\tilde{\Delta}_C$ are the (fluctuating) detunings of the pump, signal, and idler fields, respectively, and $A_\mathrm{in}$ is the pump field in the bus waveguide normalized to photon flux. All the other parameters follow the definition in section \ref{sec_STL}. We can similarly decompose the fields into their mean and fluctuating parts as,
\begin{align}
	\tilde{O} = (O+\tilde{u}_o+i\tilde{v}_o)e^{i\phi_o},
\end{align}
where $O\in\{A,B,C\}$. $A$, $B$, and $C$ correspond to the average amplitude of the fields, $\tilde{u}_a/A$, $\tilde{u}_b/B$, and $\tilde{u}_c/C$ correspond to the relative amplitude noise, and $\tilde{v}_a/A$, $\tilde{v}_b/B$, and $\tilde{v}_c/C$ correspond to the phase noise.
In addtiona, we let $\tilde{A}_\mathrm{in} = (A_\mathrm{in}+\tilde{u}_\mathrm{in}+i\tilde{v}_\mathrm{in})$ and $\tilde{\Delta}_{A,B,C} = \Delta_{A,B,C}+\tilde{\delta}_{a,b,c}$. The mean values follow the same equations as Eq. (\ref{eq_OPO_Mean1})-(\ref{eq_OPO_Mean3}). Using Eq. (\ref{eq_OPO_Mean_Property1})-(\ref{eq_OPO_Mean_Property3}) and with some calculation, we can find the difference of signal idler fluctuations as,
\begin{align}
	&\frac{d}{dt}(\tilde{u}_b-\tilde{u}_c) = -\alpha (\tilde{u}_b-\tilde{u}_c),\label{eq_Classical_Amplitude}\\
	&\frac{d}{dt}\left(\tilde{v}_b-\tilde{v}_c\right) = -(\tilde{\delta}_bB-\tilde{\delta}_cC)-2\left(BC+A^2\cos\phi\right)(\tilde{u}_b-\tilde{u}_c) ,\label{eq_Classical_Phase}
\end{align}
where $\phi = 2\phi_A-\phi_B-\phi_C$, and we have used Eqs. (\ref{eq_OPO_Mean_Property1}) - (\ref{eq_OPO_Mean_Property4}). Equation (\ref{eq_Classical_Amplitude}) indicates $\tilde{u}_b = \tilde{u}_c$ as the amplitude damps without a driving force, which physically corresponds to energy conservation. Thus, using Eq. (\ref{eq_Classical_Phase}), we get the phase difference as,
\begin{align}
	\frac{d}{dt}(\tilde{\phi}_b-\tilde{\phi}_c) = -(\tilde{\delta}_b-\tilde{\delta}_c).\label{eq_Classical_phase2}
\end{align}

Equations (\ref{eq_OPO_Classical1}) - (\ref{eq_OPO_Classical3}) can also be directly simulated. To generate Fig. 2A in the main text, we use the same cavity parameters as those in Fig. \ref{figOPOTheroy}. We assume the detuning noise follows $\tilde{\delta}_x = k_x\tilde{t}$, where $x \in \{A,B,C\}$, $k_A$, $k_B$, $k_C$ are constant coefficients, and $\tilde{t}$ is the temperature fluctuation identical to $\Delta T$ in the main text. In addition, we let 
\begin{align}
	k_A : k_B : k_C = \omega_A : \omega_B : \omega_C,\label{eq_Thermal_Ratio}
\end{align}
and characterize $k_A\tilde{t}$ using the method in section \ref{sec_Thermal_Characterize}. 

To verify Eq. (\ref{eq_Classical_phase2}), we directly simulate the stochastic field involution using Eqs. (\ref{eq_OPO_Classical1}) - (\ref{eq_OPO_Classical3}). The cavity condition and pump power are identical to those in section \ref{sec_STL}. We set the intensity noise of the pump at the shot-noise level. The phase noise of the pump is a combination of an STL-limited process with a 2-kHz-linewidth and a noise peak at 22-kHz offset frequency. The detuning fluctuations are created piecewise to resemble the experimental characterization using an approach similar to \cite{Timmer_Misc_1995}. In addition, the pump, signal, and idler detuning fluctuations are correlated according to Eq. (\ref{eq_Thermal_Ratio}). The simulation results are presented in the main article.

\section{Numerical Model of Synchronization}
The model of OPO-soliton synchronization is identical to that for the soliton-soliton synchronization presented in \cite{Jang_NatPhot_2018}, which we list here as,
\begin{align}
	&\frac{\partial E_1}{\partial t} = \left(-\frac{\alpha}{2}-i\Delta_1-i\frac{L\mathcal{F}\beta_2}{2}\frac{\partial^2}{\partial\tau^2} + i\gamma L\mathcal{F}|E_1|^2\right)E_1 + \sqrt{\kappa \mathcal{F}}E_\mathrm{in,1},\label{eq_Sync1}\\
	&\frac{\partial E_2}{\partial t} = \left(-\frac{\alpha}{2}-i\Delta_2-\tau_d\mathcal{F}\frac{\partial}{\partial\tau}-i\frac{L\mathcal{F}\beta_2}{2}\frac{\partial^2}{\partial\tau^2} + i\gamma L\mathcal{F}|E_2|^2\right)E_2 + \sqrt{\kappa \mathcal{F}}E_\mathrm{in,2} + \theta \mathcal{F}E_1,\label{eq_Sync2}
\end{align}
where $E_1$ and $E_2$ are the field envelope in the OPO and soliton-comb cavities, respectively, $\alpha$ is the cavity loss rate, $\Delta_1$ and $\Delta_2$ are the pump detunings of the OPO and soliton-comb cavities, respectively, $\tau_d$ is the difference of the roundtrip time between the cavities, $L$ is the roundtrip length, $\mathcal{F}$ is the free-spectral range (FSR), $\beta_2$ is the GVD coefficient, $\gamma$ is the nonlinear coefficient, $\kappa$ is the pump coupling rate, $\theta$ is the roundtrip coupling coefficient of the coupling link at each coupling point, $E_\mathrm{in,1}$ and $E_\mathrm{in,2}$ are the input pump fields of the OPO and soliton cavities, respectively, $t$ is the slow time on the scale of cavity lifetime, and $\tau$ is the fast time on the scale of cavity roundtrip time. Unlike the previous sections, all the fields in this section are normalized to power. We have ignored the time delay introduced by the coupling link, which can be incorporated with a frequency-dependent phase term for $\theta$.

The field evolution can be simulated using the split-step Fourier method. We use $\alpha = 2\pi\times200$ MHz, $\kappa = 2\pi\times100$ MHz, $\beta_2$ = -25 ps$^2$/km, $\mathcal{F} = 200$ GHz, and $L = 2\pi\times110 \mu$m, which is identical to those in section {\ref{sec_STL}}. We use a pump power of 20 mW and detuning of $2\pi\times50$ MHz for the OPO, and a pump of $100$ mW and a detuning of $2\pi\times800$ MHz for the soliton. To demonstrate roundtrip-time synchronization, we set a roundtrip-time difference $\tau_d = 0.02$ fs, which is shown as a drift on the fast-time grid (Fig. 2E in the main text). This drift is stopped by introducing a coupling $\theta = 0.15\%$, corresponding to a total power coupling of $\theta^2 = 2.25\times10^{-6}$ in a roundtrip time.

\begin{figure}
	\centering
	\includegraphics{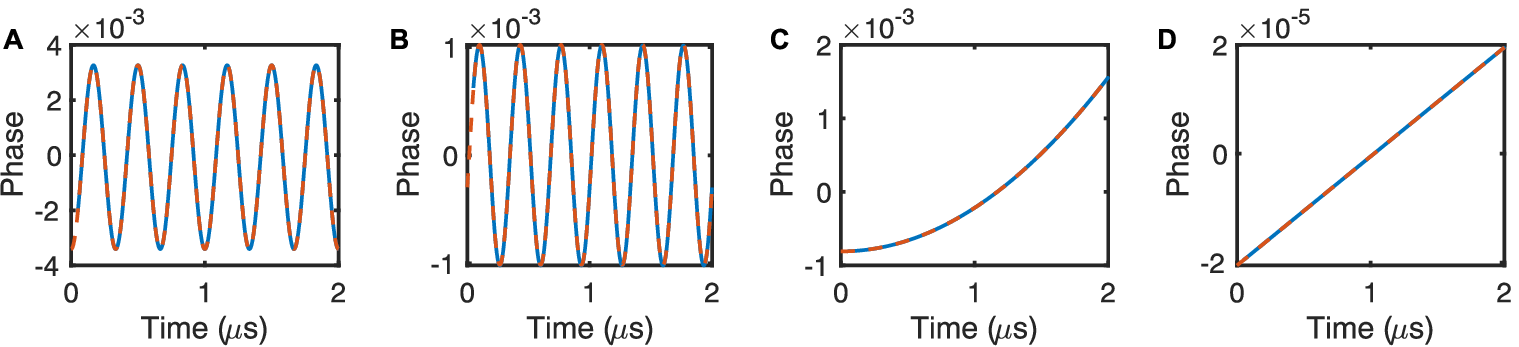}
	\caption{Characterization of noise suppression via synchronization. (A-D), The repetition-rate change induced by a sinusoidal modulation of $\tau_d$ at a frequency of 300 kHz (A, B) and 3 kHz (C, D) without the coupling link (A, C) and with the coupling link (B, D). The solid traces are numerical simulations with Eqs. (\ref{eq_Sync1}) and (\ref{eq_Sync2}), and the dashed traces are fits.}
	\label{figSynchroNoise}
\end{figure}

We simulate the suppression of soliton noise by introducing a sinusoidal modulation in $\tau_d$, which results in a sinusoidal change of soliton-repetition rate. We read out this change by extracting the phase of the comb line next to the pump, as the phase of the pump mode is constant for all $\tau_d$. We then fit the comb line phase with a sine function to extract the modulation amplitude. By comparing the amplitude with $\theta = 0$ and $\theta \neq 0$, we can get the maximum noise suppression strength at a given offset frequency. This method avoids propagation for a long time at small offset frequencies. We simulate variations at frequencies 1 kHz, 3 kHz, 10 kHz, 30 kHz, $\cdots$, 300 MHz, and 1 GHz, and show typical results in Fig. \ref{figSynchroNoise}. The noise suppression factor at an offset frequency is the ratio between the amplitudes with and without synchronization, which reflects the maximum noise-rejection capability when the OPO noise is much lower than the soliton noise. If the OPO noise is above the soliton noise minus the noise-suppression factor, the soliton noise follows the OPO noise after synchronization.

\section{Low-Thermorefractive-Noise Design}
\begin{figure}
	\centering
	\includegraphics{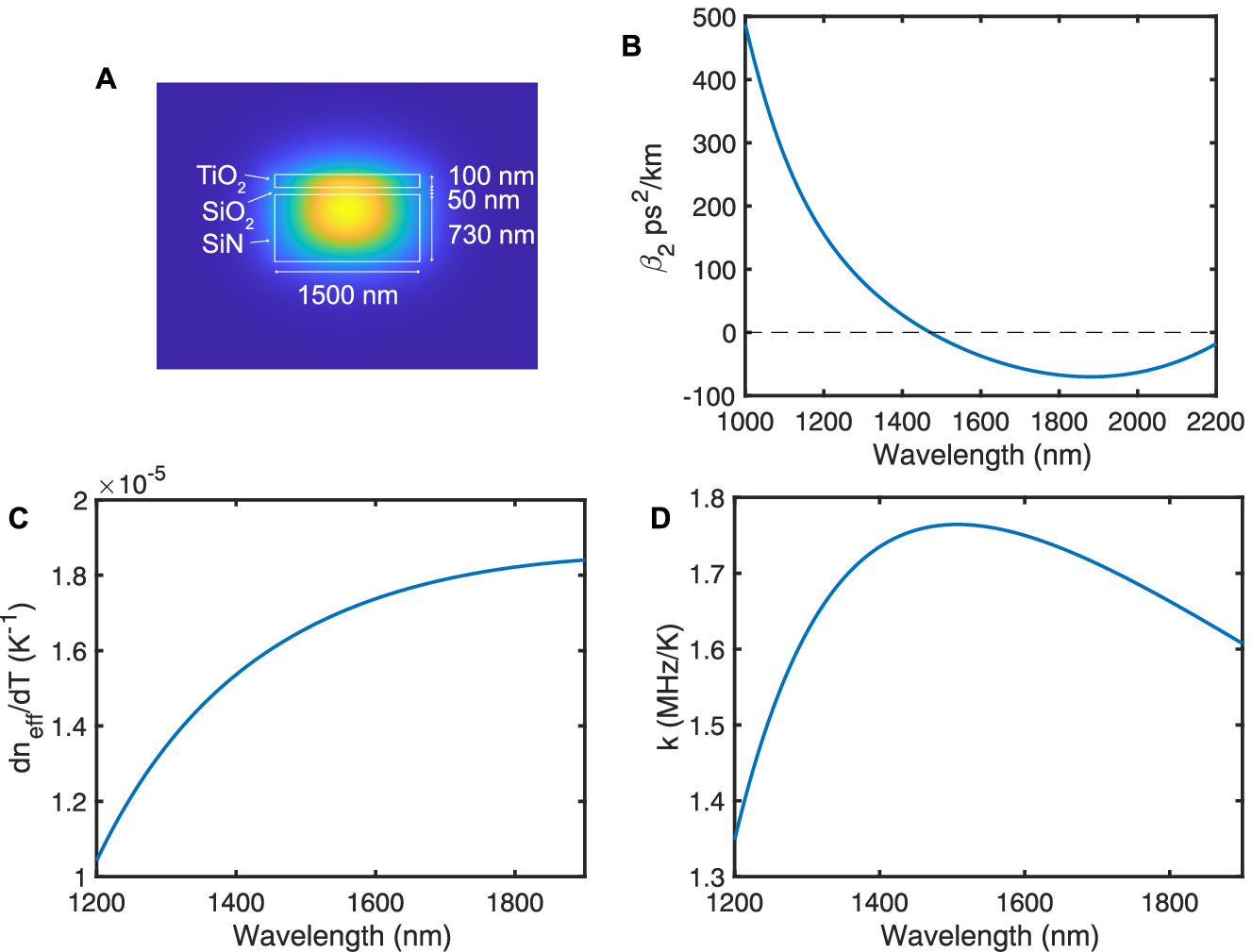}
	\caption{Design of a waveguide with tailored thermal properties for low-noise OPO. (A), the structure and simulated mode profile of the waveguide. (B), the simulated GVD of the waveguide. (C), The simulated thermorefractive coefficient of the waveguide. (D), The thermal-induced resonance shift as a function of wavelength, showing pairs of wavelengths with identical shift coefficients at the telecom wavelengths. }
	\label{figAthermal}
\end{figure}

As shown in Eq. (3), the thermorefractive noise in OPA corresponds to the mismatch between the resonance shift of the signal and idler caused by temperature fluctuations. The resonance frequency $\omega$ satisfies 
\begin{align}
\frac{n_\mathrm{eff}\omega L}{c} = 2\pi m,\label{eq_Resonance}
\end{align}
where $n_\mathrm{eff}$ is the effective index, $L$ is the cavity length, $c$ is the speed of light in vacuum, and $m$ is the resonator mode number. Differentiating Eq. (\ref{eq_Resonance}) with respect to the temperature $T$, we can get the resonance-shift coefficient,
\begin{align}
	k = \frac{d\omega}{dT} =\frac{\omega}{n_g}\frac{dn_\mathrm{eff}}{dT},
\end{align}
where $n_g$ is the group index. In a regular silicon-nitride (SiN) waveguide, $\omega$ is the dominant term, and $k$ increases as the frequency increases. However, by incorporating a small amount of TiO$_2$, which has a thermorefractive coefficient of -1$\times10^{-4}$ K$^{-1}$, we can significantly reduce $dn_\mathrm{eff}/dT$, and increase its slope as a function of wavelength. Fig. \ref{figAthermal}A shows an example of such a structure. A 100-nm thick TiO$_2$ layer is placed 50 nm away from the (SiN) core, which allows good mode confinement. We simulate the effective index at different wavelengths and temperatures using the thermorefractive coefficient for SiN as 0.4$\times10^{-4}$ K$^{-1}$ and SiO$_2$ as 0.1$\times10^{-4}$ K$^{-1}$. As shown in Fig. \ref{figAthermal}B, the waveguide exhibit anomalous GVD around 1550 nm. The effective index has a non-zero themorefractive coefficient which increases as a function of wavelength. This allows pairs of wavelengths with identical $k$ coefficients, which can be used to suppress the thermal noise in OPO operation.

\section{Thermorefractive Noise Characterization}\label{sec_Thermal_Characterize}
\begin{figure}
\centering
\includegraphics{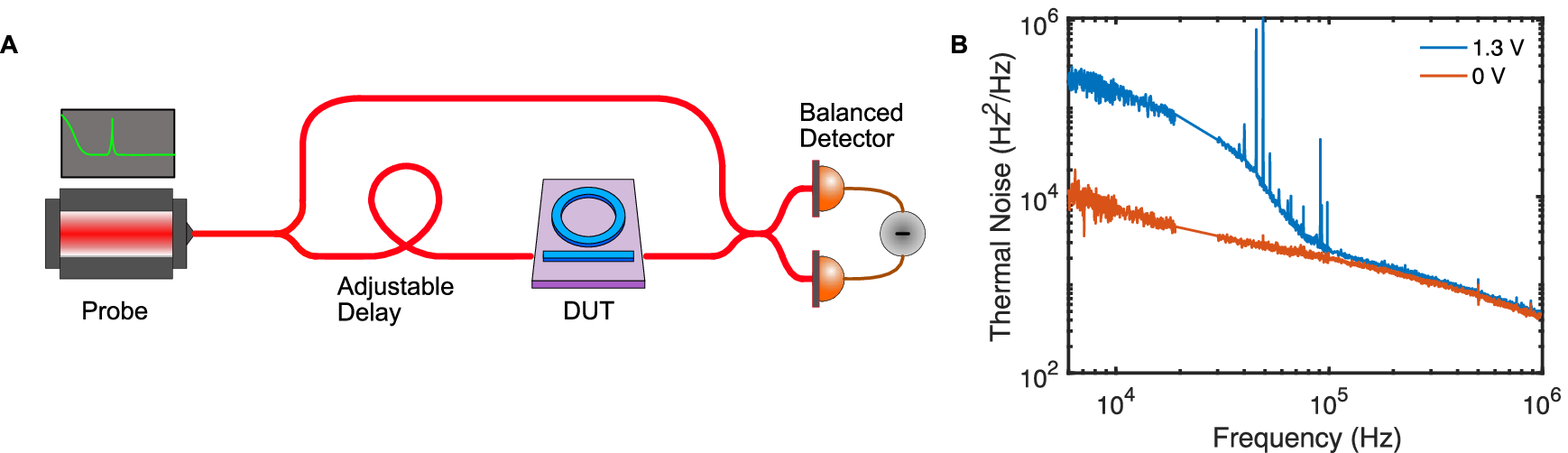}
\caption{Thermal noise characterization. A, Homodyne setup for thermal noise characterization of microresonators. DUT, device under test. B, Measured thermal noise of the SiN device at room temperature (0V) and when a heating voltage is applied using a commercial arbitrary-waveform generator (1.3 V).}
\label{figHomodyne}
\end{figure}
We characterize the cavity resonance fluctuation using a homodyne system \cite{Gorodetksy_OE_2010}. As shown in Fig. \ref{figHomodyne}A, The system consists of an MZI and a balanced photodetector. We use an adjustable delay to balance the arm lengths of the MZI to an accuracy of a few micrometers. We also attenuate the power after the adjustable delay to avoid nonlinear phase shift and heating in the device under test (DUT). The MZI is biased such that the powers are balanced on the detector. When the probe beam is in resonance with the microresonator, the intracavity field can be modeled by,
\begin{align}
	\frac{d\tilde{A}}{dt} = -\frac{\alpha}{2}\tilde{A} - i\tilde{\delta}\tilde{A} + \sqrt{\kappa}\tilde{A}_\mathrm{in},
\end{align}
where $\tilde{A}$ is the intracavity field, $\alpha$ is the loss rate, $\tilde{\delta}$ is the resonance fluctuation, $\kappa$ is the coupling rate, and $\tilde{A}_\mathrm{in}$ is the input field. We can decompose $\tilde{A}_\mathrm{in}$ and $\tilde{A}$ into their mean and fluctuations as,
\begin{align}
	&\tilde{A}_\mathrm{in} = (A_\mathrm{in}+\tilde{a}_\mathrm{in})e^{i\tilde{\psi}_\mathrm{in}},\\
	&\tilde{A} = (\frac{2\sqrt{\kappa}}{\alpha}A_\mathrm{in}+\tilde{a})e^{i\tilde{\psi}}.
\end{align}
Up to the first order in small fluctuations, $\tilde{a}$ and $\tilde{\psi}$ follow,
\begin{align}
	&\frac{d\tilde{a}}{dt} = -\frac{\alpha}{2}\tilde{a} + \sqrt{\kappa} \tilde{a}_\mathrm{in},\label{eq_Homo1}\\
	&\frac{d\tilde{\psi}}{dt} = -\frac{\alpha}{2}\tilde{\psi} - \tilde{\delta} + \frac{\alpha}{2}\tilde{\psi}_\mathrm{in}.\label{eq_Homo2}
\end{align}
Equations (\ref{eq_Homo1}) and (\ref{eq_Homo2}) can be solved in the frequency domain, which yields,
\begin{align}
	&\tilde{a}(\omega) = \frac{2\sqrt{\kappa}}{\alpha-i2\omega}\tilde{a}_\mathrm{in},\\
	&\tilde{\psi}(\omega) = -\frac{2}{\alpha-i2\omega}\tilde{\delta}+\frac{\alpha}{\alpha-i2\omega}\tilde{\psi}_\mathrm{in}.
\end{align}
The transmitted field has the form of $\tilde{A}_\mathrm{in}-\sqrt{\kappa}\tilde{A}$ which has a phase term,
\begin{align}
	\tilde{\psi}_\mathrm{out} = \frac{\alpha}{\alpha-2\kappa}\tilde{\psi}_\mathrm{in}-\frac{2\kappa}{\alpha-2\kappa}\tilde{\psi}.
\end{align}
The homodyne output voltage is proportional to the phase difference of the two arms, which has the form,
\begin{align}
	\tilde{V} \propto \tilde{\psi}_\mathrm{in}-\tilde{\psi}_\mathrm{out} = -\frac{4}{(\alpha-2\kappa)(\alpha-i2\omega)}\tilde{\delta} - \frac{4}{(\alpha-2\kappa)(\alpha-i2\omega)}\omega\tilde{\psi}_\mathrm{in}.
\end{align}
Thus, the power spectrum of $\tilde{V}$ is,
\begin{align}
	\mathcal{L}_{V}(f) \propto \mathcal{L}_{\delta}(f) + \mathcal{L}_{\omega}(f),
\end{align}
where $\mathcal{L}_{\delta}$ is the frequency noise spectrum of the resonance frequency and $\mathcal{L}_\omega$ is the frequency noise of the laser. This allows us to accurately calibrate the proportionality factor between $\mathcal{L}_V$ and $\mathcal{L}_\delta$ by introducing a high-noise tone in the probe laser. This tone can be accurately measured using the heterodyne scheme in Fig. 1D, and all other frequency components in $\mathcal{L}_V$ can be calibrated with this tone.
Figure \ref{figHomodyne}B shows the measured thermorefractive noise of the 227-GHz silicon nitride ring, which agrees with previous experiments \cite{Huang_PRA_2019}. In our experiment, a 1.3-V voltage is applied to the soliton ring which is required for matching the resonances of the two rings to the same pump. The commerical arbitrary-waveform generated used to supply this voltage adds additional noise as shown in the blue trace.

%

\end{document}